\documentclass[aps,pra,twocolumn,groupedaddress,amsmath,amssymb]{revtex4-1}
\usepackage{physics}
\usepackage{amsmath}
\usepackage{hyperref}
\usepackage{dcolumn}
\usepackage{bm}
\usepackage[pdftex]{graphicx}
\usepackage{graphicx}
\usepackage{braket}
\usepackage{amssymb}
\usepackage[utf8]{inputenc}
\usepackage[english]{babel}
\usepackage{array}

\begin{document}
\title{Quantum radiation from a shaken two-level atom in vacuum}
\author{Lezhi Lo and C. K. Law}
\affiliation{
Department of Physics and Institute of Theoretical Physics, 
The Chinese University of Hong Kong, Shatin, Hong Kong Special Administrative Region, China}
\date{\today}
\begin{abstract}
We present a non-relativistic theory of quantum radiation generated by shaking a two-level
atom in vacuum. Such radiation has the same origin of photon emission in dynamical Casimir effect. 
By performing a time-dependent ``dressing''  transformation to the Hamiltonian, we derive an 
interaction term that governs the radiation. In particular, we show that photon pairs can
be generated, not only by shaking the position of the atom, but also by changing the 
internal states of the atom. As applications of our theory, we calculate the emission
rate from an oscillating atom, and the multi-photon state generated in a single-photon 
scattering process.

 \end{abstract}
\maketitle

\section{Introduction}

In quantum mechanics, fluctuations in vacuum fields can result in a variety of observable 
physical phenomena \cite{milonni}. An interesting example is the Dynamical Casimir Effect(DCE)\cite{dodonov,DCEreview1} 
which converts vacuum fluctuations into radiation by modulating 
the system with time dependent parameters. Traditionally, DCE is studied in macroscopic 
systems such as a moving mirror \cite{fulling} or a cavity with a time-varying length \cite{moore}, 
and DCE by modulation of boundary conditions was observed in superconducting circuits \cite{nori}. 
At the microscopic level, DCE occurs when an atom moves non-uniformly in vacuum \cite{mdce}, 
and such a problem has also been discussed in the context of Unruh radiation \cite{lin1,lin2}.
Apart from moving an atom, we note that a change of internal states of a rest atom may 
also distort the vacuum non-adiabatically and emit photons \cite{Stassi1}. This is 
understood because the vacuum field can interact differently with different electronic states. 

In this paper we present a microscopic Hamiltonian which governs the generation of photons 
when an atom is subjected to time-dependent changes in its external or internal 
states. This would provide a microscopic picture of DCE and other similar parametric amplification 
processes of the quantum vacuum. Mathematically, the presence of counter-rotating terms in atom-field 
interactions is responsible for the radiation. In stationary systems these counter-rotating terms 
determine how an atom is dressed by virtual photons, and a useful technique of 
handling counter-rotating terms is the `dressing' transformation \cite{Silbey}. Such a transformation 
can significantly simplify the description because the Hamiltonian is represented in a suitable photon-atom 
dressed basis, in a way that virtual transitions between dressed states appear only as higher order 
processes. The transformation has been applied to the studies of quantum Rabi model \cite{Gan,law2}, 
spin-boson model \cite{Zheng1,Zheng2,Shi}, effects of counter-rotating terms on spontaneous decay 
\cite{Zubairy2,zhu1,zhu2}, control of Lamb shift \cite{Lamb1}, and quantum Zeno and anti-Zeno 
effects \cite{Zubairy,ZhuAQZE,Cao}. Here we generalize this transformation to time-dependent systems 
and discover an interaction term directly connected to DCE or Unruh radiation. By treating this 
term as a perturbation, we employ time-dependent perturbation theory to calculate the two-photon 
emission rate from an oscillating atom and the three-photon amplitude generated in a single-photon 
scattering process. The multiphoton state in the latter serves as a basic example of quantum 
radiation triggered by a change of internal states during a quantum process.

\section{The model Hamiltonian}
We begin with a Hamiltonian of a two-level atom interacting with a quantized electromagnetic 
field:
\begin{equation}
H=\frac{\omega_{e}}{2}\sigma_{z}+\sum_{k}\omega_{k}a_{k}^{\dagger}a_{k}+\sum_{k}[g_{k}^{*}(t)a_{k}^{\dagger}
+g_{k}(t)a_{k}]\sigma_{x},
\label{std H}
\end{equation}
where $\sigma_z = |e \rangle \langle e| - |g \rangle \langle g|$ and 
$\sigma_x = |e \rangle \langle g| + |g \rangle \langle e|$ 
are Pauli matrices describing the two-level atom with an excited state $|e\rangle$ and 
a ground state $|g\rangle$. The atomic transition frequency is denoted as $\omega_e$, and $a_k$ 
and $a_k^\dag$ are annihilation and creation operators associated with the field mode $k$ of 
frequency $\omega_k$. Note that the mode index $k$ used here is a general label for a normal mode 
of the field. For example in free space, $k$ corresponds to a wave vector ${\bf k}$.
The time-dependence of coupling strengths $g_{k}(t)$ can be realized by various settings, 
for example changing the position of the atom, and the specific expression is determined by the 
form of interaction. Throughout this paper, we assume that $g_k(t)$ changes slowly in the time scale of 
$\omega_e^{-1}$.

\subsection{Time-dependent dressing transformation}

We consider a time-dependent unitary operator defined by:
\begin{equation}
T(t)\equiv \exp[\sigma_{x}X(t)],
\label{transf def}
\end{equation}
where
\begin{equation}
X(t)\equiv \sum_{k}[\xi^{*}_{k}(t)a_{k}^{\dagger}-\xi_{k}(t)a_{k}]
\label{transformation}
\end{equation}
with $\xi_{k}(t)$'s being some small time-dependent parameters to be determined later. 
Let $\ket{\psi(t)}\equiv T(t)\ket{\Psi(t)}$ be the state in the transformed frame, where 
$\ket{\Psi(t)}$ is the state in the original frame, then the evolution of $\ket{\psi(t)}$
is governed by the transformed Hamiltonian $H'=THT^{\dagger}-iT \frac {dT^{\dagger}}{dt}$.
In Appendix \ref{H calc}, we show that $H'$ up to second order in $\xi$ is given by:
\begin{eqnarray}
H'&=&\frac{\omega_{e}}{2}\sigma_{z}+\sum_{k}\omega_{k}a_{k}^{\dagger}a_{k}
\nonumber\\
&& +\sum_{k}\sigma_{+}a_{k}\big[(\omega_{e}-\omega_{k})\xi_{k}+g_{k}-i\dot{\xi}_{k}\big]+h.c.
\nonumber\\
&&+\sum_{k}\sigma_{-}a_{k}\big[(-\omega_{e}-\omega_{k})\xi_{k}+g_{k}-i\dot{\xi}_{k}\big]+h.c.
\nonumber\\
&& +\omega_{e}\Big[\sum_{k}\big(\xi_{k}^{*}a_{k}^{\dagger}-\xi_{k}a_{k}\big)\Big]^{2}\sigma_{z}+E(t)
\label{transf H exp}
\end{eqnarray}
with $\dot{\xi}_{k}=\dv{t}\xi_{k}$. The last term 
$E(t)\equiv \sum_{k}(\frac{i}{2}\xi_{k}^{*}(t)\dot{\xi}_{k}(t)-g_{k}(t)\xi_{k}^{*}(t)+c.c.
+\omega_{k}\abs{\xi_{k}(t)}^{2})$ is a time dependent real number which only contributes a 
phase to the overall state and can be ignored. 

The aim of the transformation is to eliminate counter rotating terms in the 
third line of Eq. (\ref{transf H exp}) with a suitable set of $\{\xi_{k}(t)\}$. This is done by 
setting the coefficients of the counter-rotating terms to vanish, i.e.,
\begin{equation}
(-\omega_{e}-\omega_{k})\xi_{k}(t)+g_{k}(t)-i\dot{\xi}_{k}(t)=0,
\label{CR vanish}
\end{equation}
which has the solution:
\begin{equation}
\xi_{k}(t)=\xi_{k}(0)e^{i(\omega_{k}+\omega_{e})t}-i\int_{0}^{t}dt'g_{k}(t')e^{i(\omega_{k}+\omega_{e})(t-t')}.
\label{transformed coupling no adia}
\end{equation}
We are free to choose the initial condition $\xi_{k}(0)$. If we choose 
$\xi_{k}(0)=\frac{g_{k}(0)}{\omega_{k}+\omega_{e}}$ and make use of the assumption that 
$g_{k}(t)$ varies slowly in the time scale of $\omega_e ^{-1}$, then $\xi_{k}(t)$ is approximated by
\begin{equation}
\xi_{k}(t)\approx\frac{g_{k}(t)}{\omega_{k}+\omega_{e}}.
\label{transformed coupling}
\end{equation}
The Hamiltonian $H'$ then becomes:   
\begin{eqnarray}
H'=&\frac{\omega_{e}}{2}\sigma_{z}+\sum_{k}\omega_{k}a_{k}^{\dagger}a_{k}
+\sum_{k}(\eta_{k}a_{k}\sigma_{+}+\eta_{k}^{*}a_{k}^{\dagger}\sigma_{-})
\nonumber\\
&+\frac{1}{4\omega_{e}}\sum_{j,k}(\eta_{j}^{*}a_{j}^{\dagger}
-\eta_{j}a_{j})(\eta_{k}^{*}a_{k}^{\dagger}-\eta_{k}a_{k})\sigma_{z},
\label{transf H eta2}
\end{eqnarray}
where the new co-rotating coupling strength $\eta_{k}(t)=2\omega_{e}\xi_{k}(t)$ is defined. 

The second line of Eq. (\ref{transf H eta2}) can be put into normal order, and this yields a 
time-dependent $c$-number multiplying $\sigma_{z}$ which corresponds to a time-dependent shift 
of transition frequency between the two atomic levels. To properly account for the shift in a 
natural atom in non-relativistic theory, one can impose a frequency cut-off $\omega_c$ and 
subtract the relevant self energy terms as in the standard treatment of the Lamb shift problem \cite{milonni2}. For a multi-level atom in three-dimensional free space with constant $g_k$'s, 
it has been demonstrated that the dressing transformation and the mass renormalization procedure 
can lead to a standard expression of the Lamb shift \cite{Zubairy2,Zubairy}. Here for time-dependent 
systems, since $\eta_k (t)$ follows $g_k (t)$ adiabatically according to Eq. (\ref{transformed coupling}), 
the renormalized shift can be interpreted as a generalized (time-dependent) Lamb shift.  
For convenience we shall use $\omega_e'$ to denote the shifted transition frequency of the 
atom.

Finally, the Hamiltonian reads:
\begin{equation}
H'=H_{0}+H_{1}+ \sigma_z \Gamma , 
\label{transf H norm order}
\end{equation}
where 
\begin{eqnarray}
&& H_{0} = \frac{\omega_{e}'}{2}\sigma_{z}+\sum_{k}\omega_{k}a_{k}^{\dagger}a_{k},
\\
&& H_{1}= \sum_{k}(\eta_{k}a_{k}\sigma_{+}+h.c.),
\\
&& \Gamma (t) = \frac{1}{4\omega_{e}}\sum_{j,k}(
\eta_{j}^{*}\eta_{k}^{*}a_{j}^{\dagger}a_{k}^{\dagger}
-\eta_{j}^{*}\eta_{k}a_{j}^{\dagger}a_{k} +h.c. 
) 
\end{eqnarray}
are defined.  Note that the counter-rotating terms $a_k^+ \sigma_+$ and $a_k^- \sigma_-$ have been 
eliminated without invoking any rotating wave approximation. The transformation $T$
has taken care most of the virtual transitions or dressing effects due to counter-rotating terms, 
leaving $\sigma_z \Gamma$ as a small correction. If $\sigma_z \Gamma $ can be ignored, then the 
ground state is simply $|g\rangle |0\rangle$ (where $|0\rangle$ is the vacuum state in the transformed 
frame). Such a state in the original frame reads as $T^\dag|g\rangle |0\rangle$, a dressed 
state in which the atom and virtual photons are entangled.

We point out that $\sigma_z \Gamma$ governs the generation of radiation via the pair creation 
operators $a_{j}^{\dagger}a_{k}^{\dagger}$. Such a term has often been neglected in stationary 
systems because it is second order in $g$ and off-resonance. However, when the atom is subjected 
to time-dependent modulation, $\sigma_z \Gamma$ could lead to a resonant generation of photons. 
As a remark, we note that our Hamiltonian is different from the one derived from the atomic 
polarizability approach \cite{mdce}. The theoretical framework provided here allows us to study 
the quantum radiation process in dressed basis, and by keeping track of the internal degrees of freedom, 
DCE due to time-dependent perturbation of internal states can be addressed.

\subsection{Ground state at $t=0$}
Assuming the coupling strengths $g_k(t)$ start changing only for positive times $t>0$, the ground 
state defined at $t=0$ can serve as an initial state to study the quantum dynamics. It should be 
noted that $|g\rangle |0\rangle$ mentioned above is not the true ground state of the system because 
of the presence of the $\sigma_z \Gamma$ term. If we take $|g\rangle |0\rangle$ as an initial state, 
there will be additional radiative effects due to self-dressing of the system \cite{PassanteRadiativeSelfDressing,ValentiRadiativeSelfDressing}, 
which would obscure the quantum radiation we are interested in.
 
We construct the ground state $|\phi_0 \rangle$ of $H'$ approximately as
\begin{equation}
|\phi_0 \rangle \approx |g \rangle |0'\rangle, 
\label{gs}
\end{equation}
where $|0'\rangle$ is the lowest state of the following quadratic field Hamiltonian $H_B$:
\begin{equation}
H_B \equiv \sum_{k}\omega_{k}a_{k}^{\dagger}a_{k}-\Gamma (0). 
\label{bogo part}
\end{equation}
By perturbation theory up to first order in $\Gamma$, we have
\begin{equation}
|0' \rangle \approx   |0\rangle +\sum_{kk'}\frac{\Lambda_{kk'}(0)}{\omega_{k}+\omega_{k'}}
a_{k}^{\dagger}a_{k'}^{\dagger}|0\rangle
\label{gs field}
\end{equation}
where 
\begin{equation}
\Lambda_{kk'}(t)=\frac{\eta_{k}^{*}(t)\eta_{k'}^{*}(t)}{4\omega_{e}'(t)}
\label{dressing coeff}
\end{equation}
is defined. Note that $|g \rangle |0'\rangle$ is the ground state of  
$H_0+\sigma_z \Gamma (0)$, and it can be used to approximate the ground state of the full 
Hamiltonian $H'=H_0+\sigma_z \Gamma (0)+H_1$ at $t=0$ because $H_1$ would bring higher order 
corrections only. 

It is worth noting that although $|\phi_0 \rangle$ contains some photon pairs, they are 
virtual photons not contributing to radiation. This is understood because $|\phi_0 \rangle$ is a 
photon-atom bound state, and the corresponding photon density is localized around 
the atom as a part of the dressing.

\section{Quantum radiation}

In this section we treat $\sigma_z \Gamma$ as a weak perturbation and examine the evolution 
of the system. By first order time-dependent perturbation theory, the system state 
$\ket{\psi(t)}$ in Schrodinger picture is given by:
\begin{eqnarray}
\ket{\psi(t)} \approx U(t)|\psi (0) \rangle-i\int_{0}^{t}d\tau U(t-\tau)\sigma_{z}\Gamma(\tau)U(\tau) |\psi(0)\rangle, \nonumber \\
\label{TDPT state}
\end{eqnarray}
where $U(t)$ is the evolution operator generated by $H_{0}+H_{1}$, and the integral
containing $\Gamma(\tau)$ determines the amplitude of photons generated during 
the evolution. We point out that although the emitted photons described by $\ket{\psi(t)}$ 
are defined in the transformed frame, they are also photons in the original 
frame. This is because $\Gamma(\tau)$ is already second order in $\xi$.
As long as we keep the accuracy to this order consistently, the inverse transformation $T^\dag$ 
should be taken as an identity operator when operating on the second term in Eq. (\ref{TDPT state}).

In the following we examine two cases of photon production. The photons generated in these two cases can be considered as quantum radiation because they originate from non-adiabatic perturbations to the quantum vacuum as in the photon emission in DCE or Unruh radiation. To facilitate the calculation, 
$\eta_k$ is assumed to be a broad function of frequency as the ones given in Eq.(\ref{eta3d}) 
and Eq. (\ref{eta1d}). In addition, since the Lamb shifts are typically a tiny fraction 
of $\omega_e$, we shall approximate $\omega_e' \approx \omega_e$ as a constant. In this way we 
can write $U(t) \approx e^{-i(H_0+H_1)t}$ with $\omega_e'$ replaced by $\omega_e$ in $H_0$.

\subsection{Shaking the atom's position} 
In this case the initial state is assumed to be the ground state $|\phi_0 \rangle$ obtained 
in Eq. (\ref{gs}), and the atom is shaken so that its position ${\bf r}_A (t)$ is a function of time. 
This leads to a time-dependent coupling $g_k (t)$, whose explicit form in three-dimensional 
free space is given in Appendix \ref{rontgen form}. 
By using Eq. (\ref{gs}) and (\ref{gs field}), and keeping terms to 
first order in $\Gamma$, Eq. (\ref{TDPT state}) becomes,
\begin{eqnarray}
\ket{\psi(t)} \approx U(t)|\phi_0 \rangle+i\int_{0}^{t}d\tau U(t-\tau) 
\Gamma(\tau)|g \rangle |0\rangle.
\label{TDPT2}
\end{eqnarray}
Note that we have replaced $|\phi_0\rangle$ by $|g \rangle |0\rangle$ in the second term 
because the two-photon part in Eq. (\ref{gs field}) is first order in $\Gamma$. 
In addition, we have used $\sigma_z U(t)|g\rangle |0\rangle = -|g\rangle |0\rangle$. 

A further approximation can be made by observing that $H_{1}$ has little effect on the 
photons in the dressed ground state $|\phi_0 \rangle$. This is because real transitions described by
$H_{1}$ are only significant for photons at frequencies within several line-widths around the atomic 
transition frequency. However, we note that due to our assumption of $\eta_{k}$, photon pairs in 
$|\phi_0 \rangle$ spread out very broadly in frequency space (over many line widths), and the 
fraction of near resonance photons is very small. 
Consequently, we can write
$U(t)|\phi_0 \rangle \approx U_0 (t) |\phi_0 \rangle $, where $U_{0}(t)\equiv e^{-iH_0 t}$ is the
free evolution operator. 
The same argument can also be applied to the integrand of Eq. (\ref{TDPT2}), where 
most of the photons generated by $\Gamma$ are of the same far off-resonance nature, so that 
it is justified to make the approximation: 
$U(t-\tau)\Gamma(\tau)|g,0 \rangle \approx U_0 (t-\tau) \Gamma(\tau)|g,0 \rangle$. 
With these approximations, the perturbed state becomes:
\begin{eqnarray}
\ket{\psi(t)}&\approx & U_{0}(t)|\phi_0 \rangle
-i\int_{0}^{t}d\tau U_{0}(t-\tau)\Gamma(\tau)|g,0\rangle \nonumber \\
&=&  |g \rangle |0\rangle  
+ |g\rangle \sum_{kk'} C_{kk'} (t)a_{k}^{\dagger}a_{k'}^{\dagger} |0\rangle,
\end{eqnarray}
where
\begin{eqnarray}
C_{kk'} (t)&=& \frac{\Lambda_{kk'}(0)}{\omega_{k}+\omega_{k'}}e^{-i(\omega_{k}+\omega_{k'})t} \nonumber \\
&& + i\int_{0}^{t}d\tau
\Lambda_{kk'}(\tau) e^{-i(\omega_{k}+\omega_{k'})(t-\tau)}.
\label{time dep pert 1}
\end{eqnarray}
If the couplings, and hence $\Lambda_{kk'}$, are time independent, then the freely propagating terms in the first line of Eq. (\ref{time dep pert 1}) due to $U_{0}(t)\ket{\phi_{0}}$ will be exactly 
cancelled by the lower limit of the time integral in the second line, and $C_{kk'}$ is simply the 
dressing given in Eq. (\ref{gs field}) without producing any freely propagating photons.

However, if the couplings are time dependent, then the propagating terms are no longer cancelled, 
resulting in DCE or Unruh type radiation. As an example, consider the following coupling:
\begin{align}
\eta_{k}(t)=\eta_{k}^{0}+ik_{m}r_{m}(e^{i\omega_{m}t}\eta_{k}^{+}+e^{-i\omega_{m}t}\eta_{k}^{-}),
\label{coupling form}
\end{align}
where $k_{m}$, $r_{m}$, $\eta_{k}^{0}$, and $\eta_{k}^{\pm}$ are real, time independent numbers. 
In Appendix \ref{rontgen form} we show that this is the form of coupling taken by an electromagnetic 
field interacting with a two-level atom moving in an externally prescribed non-relativistic simple 
harmonic motion, with frequency $\omega_{m}=ck_{m}$ and amplitude $r_m$ under the long wavelength 
approximation $k_{m}r_{m}\ll 1$. Taking the continuum limit, the coupling $\eta_k (t)$ in 
Eq. (\ref{coupling form}) leads to a two-photon emission rate given by Fermi's golden rule:
\begin{align}
{\cal R}=&\frac{\pi(k_{m}r_{m})^{2}}{4\omega_{e}^{2}} \int d^{D}k \int d^{D}k'\rho(k) \rho(k')
\nonumber\\
& (\eta_{k}^{+}\eta_{k'}^{0}+\eta_{k'}^{+}\eta_{k}^{0})^{2}\delta(\omega_{k}+\omega_{k'}-\omega_{m}),
\end{align}
where $D$ is the dimension of $k$-space and $\rho(k)$ is the corresponding density of states 
of the field modes. The sum of frequencies of the emitted photon pairs concentrates at $\omega_m$.

Specializing to $D=3$ free space and using the definitions of $\eta_{k}^{0}$ and $\eta_{k}^{\pm}$ in 
Appendix \ref{rontgen form}, we find that the emission rate is given by:
\begin{equation}
{\cal R} \approx C (k_{m}r_{m})^{2}(\frac{\gamma }{\omega_{e}})(\frac{\omega_{m}}{\omega_{e}})^{7} \gamma,
\label{SHO rate}
\end{equation}
where $\gamma$ is the spontaneous decay rate of the atom and $C$ is a proportionality 
constant of order about $10^{-2}$ \cite{remark}. The scaling dependence of the system 
parameters in Eq. (\ref{SHO rate}) was also found in \cite{mdce} with a different approach.

By Eq. (\ref{SHO rate}), we note that ${\cal R}$ is extremely small because $\omega_m$ 
and $\gamma$ are much lower than $\omega_e$ in general. In particular, since emitted photons 
are limited to frequencies below $\omega_m$, and the density of state scales as $\rho(k) \propto k^2$ 
(for $D=3$), the emission rate is strongly suppressed when $\omega_m$ is low. If we consider the setup in $D=1$ 
space where $\rho(k)$ is uniform, for example in a one-dimensional waveguide, then ${\cal R}$  
will scale as $(\frac{\omega_{m}}{\omega_{e}})^{3}$ instead of $(\frac{\omega_{m}}{\omega_{e}})^{7}$ 
above.

\subsection{Shaking the atom's internal state} 

In this case, $g_k$'s are time independent but the atom is shaken internally
such that there is a time-dependent population difference between the two atomic levels. 
This would generate radiation through $\sigma_z$ in the interaction term 
$\sigma_z \Gamma$. Such radiation should be distinguished from the usual dipole radiation 
interaction, because the latter is governed by atomic coherence $\sigma_+$ or $\sigma_-$ 
instead of $\sigma_z$. A simple way of changing the atomic population is by absorption and 
re-emission of a photon. Here we show that a single photon scattering is always accompanied 
by emission of multiple photons. 

For simplicity, we consider the scattering in a one-dimensional waveguide of length $L$ 
(which will be taken to infinity in the continuum limit) and cross section area $A$, 
in which the normal modes are labeled by $k$. A positive (negative) $k$ corresponds to 
a right (left) propagating mode of frequency $\omega_k = c|k|$. Assuming $x=0$ is the 
position of the atom, the coupling is given by:
\begin{equation}
\eta_{k} = \frac{2 \omega_{e}}{\omega_{e}+\omega_{k}}\sqrt{\frac{\omega_{k}}{2\epsilon_{0}\hbar AL}}d,
\label{eta1d}
\end{equation}
where $d$ is the electric dipole matrix element. We have suppressed the polarization label of the 
field modes because the dipole is assumed to be in parallel with one of the orthogonal 
polarizations.  

At $t=0$, there is an incident single photon wavepacket at a 
far distance from the atom, such that the atom prepared in the dressed ground 
state $\ket{\phi_{0}}$ would not experience the incident photon initially. The initial 
state of the system in the transformed frame is given by,
\begin{align}
\ket{\psi(0)}=W^{\dagger} \ket{\phi_{0}}
\label{IC1}
\end{align}
where $W^\dag$ is a creation operator of the single photon wavepacket defined by
\begin{align}
W^{\dagger}=\sum_{k}W_{k}a_{k}^{\dagger}.
\label{IC2}
\end{align}
Here $W_k$ are coefficients determining the shape of the wavepacket. Noting that the dressing 
transformation $T$ only modifies the field in the neighborhood of the atom, the transformation does 
not affect the initial photon as long as the wavepacket is sufficiently far away from the atom. 
Mathematically, this corresponds to the condition $[W^\dag, T]=[W^\dag, T^\dag] =0$, so that 
the initial state in the original frame is: 
$T^\dag \ket{\psi(0)}= T^\dag W^{\dagger} \ket{\phi_{0}}= W^{\dagger} T^\dag \ket{\phi_{0}}$. 

Specifically, let us consider the following Lorentizian photon wavepacket defined by:
\begin{align}
W_{k}=\sqrt{\frac{\gamma'}{cL}}\frac{e^{-i(k-k_{e})x_{0}}}{-i(k-k_{e})+\frac{\gamma'}{2c}},
\label{IC3}
\end{align}
where $x=x_{0}<0$ is the position of the front edge of the packet on the left of the atom and 
$\gamma' \ll \omega_e$ is a positive real number characterizing the spectral width of the packet. 
In addition, we consider $ck_{e}=\omega_{e}$ so that the incident photon is in resonance 
with the atom and travels to the right. Note that $W^\dag$ commutes with $T$ and $T^\dag$ 
as $\abs{x_{0}} \to \infty$.

The perturbed state given by Eq. (\ref{TDPT state}) can be evaluated approximately. 
Together with the incident photon, there can be three freely propagating photons in the final state after 
the scattering is completed. The calculation is presented in Appendix \ref{3 photon}, we find 
that the three-photon amplitude in the long time limit is approximately given by:
\begin{equation}
\ket{\psi_{3}}\approx \sum_{jkl}C_{jkl}e^{-i(\omega_{jkl}-\frac{\omega_{e}}{2})
(t-t_{0})}a_{j}^{\dagger}a_{k}^{\dagger}a_{l}^{\dagger}\ket{0},
\label{3p state}
\end{equation}
where
\begin{equation}
C_{jkl}= \frac{\sqrt{\gamma'\gamma}}{2\omega_{e}}\frac{\eta_{j}^{*}\eta_{k}^{*}
\eta_{l}^{*}}{(i\Delta_{l}-\frac{\gamma}{2})(i\Delta_{jkl}-\frac{\gamma}{2})(i\Delta_{jkl}-\frac{\gamma'}{2})}.
\label{3p coeff}
\end{equation}
Here we have defined 
$\Delta_{l}\equiv \omega_{l}-\omega_{e}$; 
$\Delta_{jkl}\equiv \omega_{j}+\omega_{k}+\omega_{l}-\omega_{e}$, 
and $t_{0}\equiv \frac{\abs{x_{0}}}{c}$ is the time needed for the photon wave packet travel 
from its initial position to the atom's. 

Eq. (\ref{3p coeff}) indicates that $C_{jkl}$ is significant when the frequency sum of the three 
photons is near the atomic transition frequency $\omega_e$, 
i.e., $\omega_j+\omega_k+\omega_l \approx \omega_e$. Since the numerator 
$\eta_k^* \eta_j^* \eta_l^* \propto \sqrt{\omega_k \omega_j \omega_l}$, the three-photon amplitude 
on the $\omega_j+\omega_k+\omega_l \approx \omega_e$ surface is not sensitive to the single-photon
resonance associated with the  $i\Delta_{l}-\frac{\gamma}{2}$ denominator. For example, in the case 
of $\gamma\approx \gamma'$ and $\omega_{j}=\omega_{k}$,
$C_{jkl}\sim c^{3/2} \gamma^{1/2}/\omega_e^2 L^{3/2}$ over the entire range 
$\omega_{l}\in (\gamma,\omega_{e}-\gamma)$.

\section{Conclusion}

To conclude, we have developed a Hamiltonian for the study of DCE or Unruh type radiation at 
the microscopic level. Through the time-dependent dressing transformation $T$, we are able to identify 
the interaction term $\sigma_z \Gamma$ which governs photon pair 
generation when modulations are applied to atom-field couplings or the atom's internal states. 
As demonstrated by the examples in Sec. III, the radiation is extremely weak in natural systems 
because the atomic transition frequency $\omega_e$ is generally much higher than the spontaneous 
emission rate $\gamma$ and mechanical modulation frequency $\omega_m$. However, recent progress 
of realizing ultrastrong coupling in artificial systems could be an important step towards the 
observation of such radiation \cite{Solano,KockumNoriUltrastrong}. In particular, the value of $\gamma$ 
can be a signficant fraction of $\omega_e$ in waveguide QED \cite{Lupascu}. We also point out that 
related quantum radiation effects based on various modulation schemes \cite{Stassi1,Ciuti,huang,Stassi2,Kockum,KockumNoriUltrastrong} 
and photon scattering \cite{Shi,GheeraertUltraStrongWaveguide} in ultrastrong coupling regime have been 
reported recently. In the future, we hope to explore applications of the time dependent dressing 
transformation in ultrastrong coupling problems and quantum radiation with multiple atoms. 

\appendix

\section{Derivation of the transformed Hamiltonian}\label{H calc}
The derivation is similar to that in stationary systems (see for example \cite{Zheng2}), the main difference
is the time-dependence of $\xi_k (t)$ which generates extra terms in the Hamiltonian:
\begin{align}
H'=THT^{\dagger}-iT\pdv{t}T^{\dagger}.
\end{align}
By $T=e^ {\sigma_{x}X }=\cosh(X)I+\sinh(X)\sigma_{x}$, we have
\begin{align}
T\sigma_{z}T^{\dagger}=\cosh (2X) \sigma_{z} - i\sinh (2X) \sigma_{y}. \label{sinh}
\end{align}
We can also consider $T$ as a spin-dependent displacement operator, with $Ta_{k}T^{\dagger}=a_{k}-\sigma_{x}\xi_{k}^{*}$. As such,
\begin{align}
&T\big[\sum_{k}\omega_{k}a_{k}^{\dagger}a_{k}+\sum_{k}(g_{k}^{*}a_{k}^{\dagger}+g_{k}a_{k})\sigma_{x}\big]T^{\dagger}
\nonumber\\
=&\sum_{k}\omega_{k}(a_{k}^{\dagger}-\sigma_{x}\xi_{k})(a_{k}-\sigma_{x}\xi_{k}^{*})
\nonumber\\
&+\sum_{k}\big[g_{k}^{*}(a_{k}^{\dagger}-\sigma_{x}\xi_{k})+g_{k}(a_{k}-\sigma_{x}\xi_{k}^{*})\big]\sigma_{x}.
\end{align}
Next we employ the expansion,
\begin{align}
e^{S}\pdv{t} e^{-S}=-\dot{S}-\frac{1}{2}\comm{S}{\dot{S}}-\frac{1}{6}\comm{S}{\comm{S}{\dot{S}}}-...
\end{align}
which gives
\begin{align}
T\pdv{t}&T^{\dagger}=-\sigma_{x}\dot{X}-\frac{1}{2}\comm{X}{\dot{X}}-...
\nonumber\\
=&-\sigma_{x}\sum_{k}(\dot{\xi}^{*}_{k}a_{k}^{\dagger}-\dot{\xi}_{k}a_{k})-\frac{1}{2}\sum_{k}(\xi_{k}^{*}\dot{\xi_{k}}-\xi_{k}\dot{\xi_{k}}^{*})
\end{align}
where $\dot{\xi}_{k}=\dv{t}\xi_{k}$. This is exact since the second order nested commutator is a c-number, 
causing all higher order nested commutators to vanish. The transformed Hamiltonian is therefore:
\begin{align}
H'=&\sum_{k}\omega_{k}(a_{k}^{\dagger}-\sigma_{x}\xi_{k})(a_{k}-\sigma_{x}\xi_{k}^{*})
\nonumber\\
&+\sum_{k}\big[g_{k}^{*}(a_{k}^{\dagger}-\sigma_{x}\xi_{k})+g_{k}(a_{k}-\sigma_{x}\xi_{k}^{*})\big]\sigma_{x}
\nonumber\\
&+\frac{\omega_{e}}{2}\Big\{\cosh [2\sum_{k}(\xi_{k}^{*}a_{k}^{\dagger}-\xi_{k}a_{k})] \sigma_{z} 
\nonumber\\
&- i\sinh [2\sum_{k}(\xi_{k}^{*}a_{k}^{\dagger}-\xi_{k}a_{k})] \sigma_{y}\Big\}
\nonumber\\
&+i\sigma_{x}\sum_{k}(\dot{\xi}^{*}_{k}a_{k}^{\dagger}-\dot{\xi}_{k}a_{k})+\frac{i}{2}\sum_{k}(\xi_{k}^{*}\dot{\xi_{k}}-\xi_{k}\dot{\xi_{k}}^{*}).
\label{transf H exact}
\end{align}
We expand $\cosh(2X)$ and $\sinh(2X)$ in powers of $\xi$. By keeping terms up to $\xi^2$, 
we obtain the form of $H'$ given in Eq. (\ref{transf H exp}). Note that for quantum states 
near the vacuum considered in this paper, $\xi^3$ and higher power terms in the expansion 
can be neglected provided that $\sum_{k}\abs{\xi_{k}}^{2} \ll 1$.

\section{The coupling $\eta_k$ of a moving atom in free space}\label{rontgen form}
The interaction between a moving atom and the electromagnetic field takes the following form under the dipole 
approximation:
\begin{align}
H_{\text{int}}=&-\vb{d}\vdot\vb{E^{\perp}}(\vb{r}_{A})
\nonumber\\
&+\frac{1}{2m_{A}}\Big\{ \vb{p}_{A}\vdot\big[ \vb{d}\cross\vb{B}(\vb{r}_{A}) \big]+\big[ \vb{d}\cross\vb{B}(\vb{r}_{A}) \big]\vdot\vb{p}_{A} \Big\}
\label{rontgen H}
\end{align}
where $\vb{d}$ is the electric dipole of the atom; $\vb{E}^{\perp}(\vb{r}_{A})$ and $\vb{B}(\vb{r}_{A})$ are 
the transverse electric and magnetic fields at the position of the atom, $\vb{r}_{A}$; $\vb{p}_{A}$ is the 
canonical momentum of the center of mass of the atom; and $m_{A}$ is the mass of the atom. The second term is 
known as the R\"ontgen term \cite{baxer,wilkens}, which arises from the magnetic field interacting with the 
magnetic dipole moment due to the motion of the electric dipole. 

By expanding the field operators in plane-wave modes in free space, 
the interaction Hamiltonian can be written as:
\begin{equation}
H_{\text{int}}=\sum_{\vb{k},s}\big[g_{\vb{k},s}(t)a_{\vb{k},s}+g_{\vb{k},s}^{*}(t)a_{\vb{k},s}^{\dagger}\big]\sigma_{x}
\end{equation}
where 
\begin{eqnarray}
g_{\vb{k},s} &=& \chi_{k}^{}e^{i\vb{k}\vdot\vb{r}_{A}(t)}
\nonumber\\
&& \times\Big\{ \hat{d}\vdot\hat{\epsilon}_{\vb{k},s}+\bm{ \beta }(t)\vdot\big[ \hat{\epsilon}_{\vb{k},s}
(\hat{d}\vdot\hat{k})-\hat{k}(\hat{d}\vdot\hat{\epsilon}_{\vb{k},s}) \big] \Big\} \nonumber \\
\end{eqnarray}
with
\begin{eqnarray}
&& \chi_{k}^{}=\sqrt{\frac{\omega_{k}}{2\epsilon_{0}\hbar V}}d,
 \\
&& \vb{d}=\bra{e}e\vb{r}\ket{g}=d\hat{d},
\\
&& \vb{k}=k\hat{k},
\\
&& \bm{\beta}(t)=\frac{\vb{\dot{r}}_{A}(t)}{c}.
\end{eqnarray}
Here $V$ is the quantization volume, ${\epsilon}_{\vb{k},s}$ is the polarization unit vector of mode $\vb{k}$ 
with $s$ polarization. All hatted quantities are unit vectors. 

Next we consider the trajectory of the atom given by $\vb{r}_{A}(t)=r_m \cos \omega_m t~ \hat r_m $ with 
$r_m \omega_m \ll c$ in the non-relativistic regime. In addition, we assume the long wavelength condition 
$k r_m \ll 1$ for field modes that are effectively involved in the two-photon emission. This condition is 
consistent with the fact that the two photons emitted have their sum of frequencies around $\omega_m$ for 
non-relativistic motion. Consequently we take the approximation: 
$e^{i\vb{k}\vdot\vb{r}_{A}(t)} \approx 1+i\vb{k}\vdot\vb{r}_{A}(t)$, and obtain
\begin{equation}
\eta_{\vb{k},s}(t)=\eta_{\vb{k},s}^{0}+ik_{m}r_{m}(e^{i\omega_{m}t}\eta_{\vb{k},s}^{+}+e^{-i\omega_{m}t}\eta_{\vb{k},s}^{-})
\label{eta3d}
\end{equation}
where $k_m=\omega_m/c$, and we have defined the following real quantities:
\begin{eqnarray}
 \eta_{\vb{k},s}^{0} &\equiv &\frac{\chi_{k}}{1+\frac{\omega_{k}}{\omega_{e}}}2(\hat{d}\vdot\hat{\epsilon}_{\vb{k},s}),
\\
 \eta_{\vb{k},s}^{\pm} &\equiv &\frac{\chi_{k}}{1+\frac{\omega_{k}}{\omega_{e}}}\Big\{ (\frac{\vb{k}}{k_{m}}\vdot\hat{r}_{m})(\hat{d}\vdot\hat{\epsilon}_{\vb{k},s})
\nonumber\\
&& \pm\hat{r}_{m}\vdot\big[ \hat{\epsilon}_{\vb{k},s}(\hat{d}\vdot\hat{k})-\hat{k}(\hat{d}\vdot\hat{\epsilon}_{\vb{k},s}) \big] \Big\}.
\end{eqnarray}

\section{Calculation of the three-photon amplitude generated by single-photon scattering}\label{3 photon}
We start with the initial state Eq. (\ref{IC1}),
\begin{eqnarray}
\ket{\psi(0)}=(1+\sum_{kk'}\frac{\Lambda_{kk'}}{\omega_{kk'}}a_{k}^{\dagger}a_{k'}^{\dagger})\sum_{k} W_k a_{k}^{\dagger}\ket{g,0},
\end{eqnarray}
where $\Lambda_{kk'}$ are time-independent. 

The first order time-dependent state is given by the second term in Eq. (\ref{TDPT state}). To evaluate the integral,
we note that the dressed photon pairs have a very broad spectrum, so they barely interact with the atom through $U(t)$. Hence we 
can approximate their evolution under $U(t)$ as free, i.e.,
\begin{align}
U(t)\sum_{kk'}\frac{\Lambda_{kk'}}{\omega_{kk'}}a_{k}^{\dagger}a_{k'}^{\dagger}U^{\dagger}(t)\approx \sum_{kk'}\frac{\Lambda_{kk'}}{\omega_{kk'}}a_{k}^{\dagger}a_{k'}^{\dagger}e^{-i(\omega_{k}+\omega_{k'})t}.
\label{non int approx}
\end{align}
Hence for the first term in Eq. (\ref{TDPT state}), 
\begin{align}
&U(t)\ket{\psi(0)}
\nonumber\\
\approx&(1+\sum_{kk'}\frac{\Lambda_{kk'}}{\omega_{kk'}}a_{k}^{\dagger}a_{k'}^{\dagger}e^{-i(\omega_{k}+\omega_{k'})t})U(t)\sum_{k''}W_{k''}a_{k''}^{\dagger}\ket{g,0}.
\label{1st term}
\end{align}
As we shall see, the last term in Eq. (\ref{1st term}) does not contribute to three-photon emission
because its propagating part will be canceled by the second term in Eq. (\ref{TDPT state}).

For the second term in Eq. (\ref{TDPT state}), we focus on the part that corresponds to three photons in the final state,
which is:
\begin{eqnarray}
-i\int_{0}^{t}d\tau U(t-\tau)\sum_{kk'}\Lambda_{kk'}a_{k}^{\dagger}a_{k'}^{\dagger}\sigma_{z}U(\tau)W^{\dagger}|g,0\rangle, \nonumber
\end{eqnarray}
where we have dropped the dressing terms in $\ket{\psi(0)}$ since their contributions are of higher order.
Next we insert $I=U^{\dagger}(t-\tau)U(t-\tau)$ after $a_{k}^{\dagger}a_{k'}^{\dagger}$ and make approximations similar 
to Eq. (\ref{non int approx}), this gives:
\begin{align}
&-i\int_{0}^{t}d\tau U(t-\tau)\sum_{kk'}\Lambda_{kk'}a_{k}^{\dagger}a_{k'}^{\dagger}\sigma_{z}U(\tau)W^{\dagger}|g,0\rangle
\nonumber\\
\approx&+i\int_{0}^{t}d\tau \sum_{kk'}\Lambda_{kk'}a_{k}^{\dagger}a_{k'}^{\dagger}e^{-i(\omega_{k}+\omega_{k'})(t-\tau)}U(t)W^{\dagger}\ket{g,0}
\nonumber\\
&-2i\int_{0}^{t}d\tau \sum_{kk'}\Lambda_{kk'}a_{k}^{\dagger}a_{k'}^{\dagger}e^{-i(\omega_{k}+\omega_{k'})(t-\tau)}
\nonumber\\
&\hphantom{-2i\int_{0}^{t}d\tau \sum_{kk'}}U(t-\tau)\ket{e}\bra{e}U(\tau)W^{\dagger}\ket{g,0}
\label{pair production term}
\end{align}
where we have replaced $\sigma_z$ by $2|e\rangle \langle e| -1$. In the first integral of 
Eq. (\ref{pair production term}), the lower limit cancels the propagating term in Eq. (\ref{1st term}), leaving 
the upper limit as the original dressed photon pair which is bounded to the atom after the scattering. 

The second integral of Eq. (\ref{pair production term}) contains the photon pair production terms 
dependent on population in the atomic excited state. It is this integral that determines the three-photon emission.

To evaluate $U(t-\tau)\ket{e}\bra{e}U(\tau)W^{\dagger}\ket{g,0}$, we note that
\begin{eqnarray}
U(t)\ket{e}&=&e^{(-{\gamma}-i{\omega_{e}})t/2}\ket{e}
\nonumber\\
&& +\sum_{k}\frac{\eta_{k}^{*} e^{-i\omega_e t/2}}{\frac{\gamma}{2}-i\Delta_{k}}(e^{-i\Delta_k t}-e^{-\gamma t/2})\ket{g,k} \nonumber \\
\label{decay sol}
\end{eqnarray}
is the well-known solution to spontaneous atomic decay. In addition, if we choose $W_{k}$ to take the Lorentzian form given by Eq. (\ref{IC3}), then the single photon scattering excited state amplitude is:
\begin{align}
\bra{e}U(\tau)W^{\dagger}\ket{g,0}=\frac{2i\sqrt{\gamma'\gamma}}{\gamma-\gamma'}(e^{-\frac{\gamma}{2}\tau}-e^{-\frac{\gamma'}{2}\tau})e^{-i\frac{\omega_{e}}{2}\tau}.
\label{sc sol}
\end{align}
By using Eq. (\ref{decay sol}-\ref{sc sol}), and working out the second integral of Eq. (\ref{pair production term}), 
we obtain the freely propagating three-photon amplitude given by Eq. (\ref{3p state}-\ref{3p coeff}) 
in the long time limit.


\begin{thebibliography}{99}
\bibitem{remark} For the two-level atomic model used in this paper, $C$ depends on the angle between 
the electric dipole moment and the velocity of the atom, and its value varies between 0.002 to 0.02 over 
the range of angles. A more realistic calculation of the emission rate should employ a multi-level 
atom model such that different directional dependence for the dipole transition elements can be included.
For example in a hydrogen atom, the ground state can couple to three degenerate excited P states
corresponding to the three orthogonal components of the dipole moment, and hence it is better described by 
a four-level model. Our Hamiltonian can be generalized to this four-level problem, and $C$ is found to be 
$\frac{23}{1260\pi}$.
\end{thebibliography}
\end{document}